\documentclass[aps,prl,twocolumn,showpacs,preprintnumbers,amsmath,amssymb,superscriptaddress]{revtex4}
\usepackage{mathptm}  
\usepackage{dcolumn}                    
\usepackage{bm}                        
\usepackage{graphicx}
\usepackage{times}
\usepackage{epstopdf}
\begin{document}

\newcommand{\Imag}{{\Im\mathrm{m}}}   
\newcommand{\Real}{{\mathrm{Re}}}   
\newcommand{\im}{\mathrm{i}}        
\newcommand{\talpha}{\tilde{\alpha}}
\newcommand{\ve}[1]{{\mathbf{#1}}}

\newcommand{\x}{\lambda}  
\newcommand{\y}{\rho}     
\newcommand{\T}{\mathrm{T}}   
\newcommand{\Pv}{\mathcal{P}} 
\newcommand{\vk}{\ve{k}} 
\newcommand{\vp}{\ve{p}} 

\newcommand{\N}{\underline{\mathcal{N}}} 
\newcommand{\Nt}{\underline{\tilde{\mathcal{N}}}} 
\newcommand{\g}{\underline{\gamma}} 
\newcommand{\gt}{\underline{\tilde{\gamma}}} 

\newcommand{\vecr}{\ve{r}} 
\newcommand{\vq}{\ve{q}} 
\newcommand{\ca}[2][]{c_{#2}^{\vphantom{\dagger}#1}} 
\newcommand{\cc}[2][]{c_{#2}^{{\dagger}#1}}          
\newcommand{\da}[2][]{d_{#2}^{\vphantom{\dagger}#1}} 
\newcommand{\dc}[2][]{d_{#2}^{{\dagger}#1}}          
\newcommand{\ga}[2][]{\gamma_{#2}^{\vphantom{\dagger}#1}} 
\newcommand{\gc}[2][]{\gamma_{#2}^{{\dagger}#1}}          
\newcommand{\ea}[2][]{\eta_{#2}^{\vphantom{\dagger}#1}} 
\newcommand{\ec}[2][]{\eta_{#2}^{{\dagger}#1}}          
\newcommand{\su}{\uparrow}    
\newcommand{\sd}{\downarrow}  
\newcommand{\Tkp}[1]{T_{\vk\vp#1}}  
\newcommand{\muone}{\mu^{(1)}}      
\newcommand{\mutwo}{\mu^{(2)}}      
\newcommand{\epsk}{\varepsilon_\vk}
\newcommand{\epsp}{\varepsilon_\vp}
\newcommand{\e}[1]{\mathrm{e}^{#1}}
\newcommand{\dif}{\mathrm{d}} 
\newcommand{\diff}[2]{\frac{\dif #1}{\dif #2}}
\newcommand{\pdiff}[2]{\frac{\partial #1}{\partial #2}}
\newcommand{\mean}[1]{\langle#1\rangle}
\newcommand{\abs}[1]{|#1|}
\newcommand{\abss}[1]{|#1|^2}
\newcommand{\Sk}[1][\vk]{\ve{S}_{#1}}
\newcommand{\pauli}[1][\alpha\beta]{\boldsymbol{\sigma}_{#1}^{\vphantom{\dagger}}}

\newcommand{\eq}{Eq.}
\newcommand{\eqs}{Eqs.}
\newcommand{\cf}{\textit{cf. }}
\newcommand{\ie}{\textit{i.e. }}
\newcommand{\eg}{\textit{e.g. }}
\newcommand{\etal}{\emph{et al.}}
\def\i{\mathrm{i}}

\title{Unconventional superconductivity on a topological insulator}

\author{Jacob Linder}
\affiliation{Department of Physics, Norwegian University of
Science and Technology, N-7491 Trondheim, Norway}

\author{Yukio Tanaka}
\affiliation{Department of Applied Physics, Nagoya University, Nagoya, 464-8603, Japan}

\author{Takehito Yokoyama}
\affiliation{Department of Applied Physics, University of Tokyo, Tokyo 113-8656, Japan}

\author{Asle Sudb{\o}}
\affiliation{Department of Physics, Norwegian University of
Science and Technology, N-7491 Trondheim, Norway}

\author{Naoto Nagaosa}
\affiliation{Department of Applied Physics, University of Tokyo, Tokyo 113-8656, Japan}
\affiliation{Cross Correlated Materials Research Group (CMRG), ASI, RIKEN, WAKO 351-0198, Japan}

\date{\today}

\begin{abstract}

We study proximity-induced superconductivity on the surface of a topological insulator (TI), focusing on unconventional pairing. We find that the excitation spectrum becomes gapless for any spin-triplet pairing, such that both subgap bound states and Andreev reflection is strongly suppressed. For spin-singlet pairing, the zero-energy surface state in the $d_{xy}$-wave case becomes a Majorana fermion, in contrast to the situation realized in the topologically trivial high-$T_c$ cuprates. We also study the influence of a Zeeman field on the surface states. Both the magnitude and direction of this field is shown to strongly influence the transport properties, in contrast to the case without TI. We predict an experimental signature of the Majorana states via conductance spectroscopy. 

\end{abstract}
\maketitle

Topological insulators represent a new state of matter which presently is generating much interest \cite{konig_jpsj_08, kane_prl_05, bernevig_06,3d}. While being insulating in the bulk due to a charge excitation gap, spin-dependent conducting channels are formed at the edges or surfaces of such materials. These states form as Kramer pairs which are topologically protected, persisting in the presence of disorder as long as time-reversal symmetry is preserved. The allure of topological insulators stems not only from their obvious interest from a fundamental physics point of view, but also because they may find use in spintronics \cite{Yokoyama}. Recent experiments have observed the surface Dirac states characteristic for 3D topological insulators \cite{hsieh}.

Another motivation for studying topological insulators is that they provide an arena for excitations that satisfy non-Abelian statistics: so-called Majorana fermions \cite{Majoranap}. Elementary excitations with non-Abelian statistics form a centerpiece in recent proposals for topological quantum computation \cite{fu_prl_08}. Majorana fermions have been shown to exist as surface states at the junction between a superconductor (S) and ferromagnetic insulator (FI) deposited on a topological insulator due to the proximity effect \cite{fu_prl_09, akhmerov_prl_09, tanaka_prl_09}. The formation of Majorana states at the interface between a superconductor and a topological insulator \cite{law_arxiv_09, santos_arxiv_09} and at the boundary of superfluid $^3$He-B \cite{volovik_arxiv_09} has also been proposed. Very recently, an experimental study reported the observation of doping-induced superconductivity in the topological insulator (TI) Bi$_2$Se$_3$ \cite{hor_arxiv_09}.

By depositing superconducting materials with an unconventional pairing symmetry on top of a TI, an exciting prospect of an interplay between the internal phase of the superconducting order parameter $\Delta$ and Majorana states opens up. In this Letter, we investigate how spin-triplet and spin-singlet $d$-wave pairing interact with the environment of a TI. We find that spin-triplet pairing universally gives rise to gapless excitations and that both bound-states and Andreev reflection are strongly suppressed. For spin-singlet pairing, we find that the zero-energy surface states in the $d_{xy}$-wave case are now Majorana fermions in contrast to the case of the topologically trivial case of the high-$T_c$ cuprates. Several works have previously investigated zero-energy vortex core states in the context of $^{3}$He and cold atoms \cite{Volovik}. While these Majorana state are localized in the vortex core, the present Majorana fermions spread along the interface to the superconducting region. Moreover, we show how the traditional zero-bias conductance peak (ZBCP) serving as a hallmark of the $d_{xy}$-wave pairing state \cite{hu_tanaka} is substantially modified in the presence of a Zeeman field. In fact, the characteristic conductance spectra of $s$-wave and $d$-wave pairing may be interchanged in the presence of a time-reversal symmetry breaking field. Also, we find that the conductance exhibits qualitatively very distinct behavior with respect to the orientation of the field, in complete contrast to the topologically trivial case. 

\begin{figure}[t!]
\centering
\resizebox{0.30\textwidth}{!}{
\includegraphics{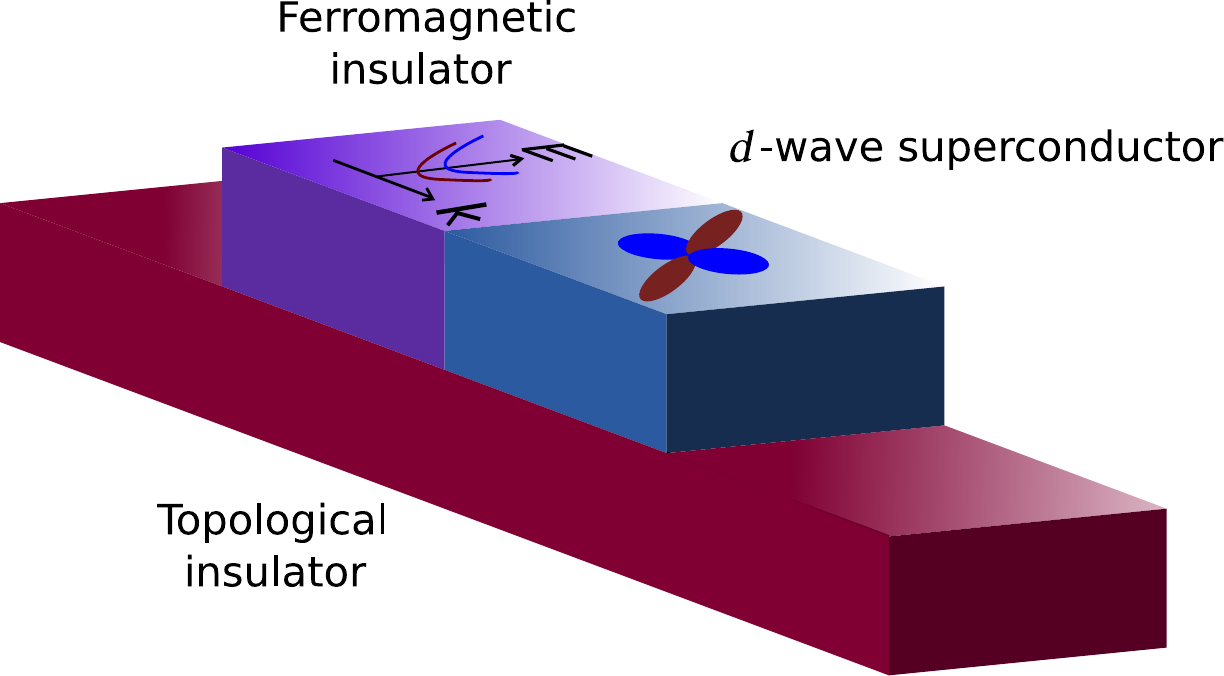}}
\caption{(Color online) We consider a TI where superconductivity and/or magnetic correlations are induced on the surface via the proximity effect to host materials with the desired properties. 
}
\label{fig:model} 
\end{figure}

We will employ a Bogolioubov-de Gennes approach to obtain the bound-states and transport properties of the system under consideration.
Using a Nambu basis $\Psi=(\psi_\uparrow, \psi_\downarrow, \psi_\uparrow^\dag, \psi_\downarrow^\dag)$, the general Hamiltonian for the surface of a TI reads:
\begin{align}\label{eq:H}
\hat{H} &= \begin{pmatrix}
\underline{H_0}(\vk) & \underline{\Delta}(\vk)\\
-\underline{\Delta}^*(-\vk) & -\underline{H_0}^*(-\vk)\\
\end{pmatrix},
\end{align}
where $\underline{H_0}(\vk) = v_F(\underline{\sigma_x}k_x +\underline{\sigma_y}k_y) - \mu$ ($\underline{\ldots}$ denotes a $2\times2$ matrix). The gap matrix $\underline{\Delta}(\vk)$ depends on both the orbital- and spin-symmetry of the Cooper pair. For a spin-singlet symmetry such as $s$-wave or $d$-wave, one finds that $\underline{\Delta}(\vk) = \Delta(\vk)\i\underline{\sigma_y}$. Diagonalization of Eq. (\ref{eq:H}) then yields the standard eigenvalues $\varepsilon = \eta\sqrt{(v_F|\vk|-\beta\mu)^2+|\Delta(\vk)|^2}$, where $\eta=\pm1$, $\beta=\pm1$. Turning to the spin-triplet case, where  $\underline{\Delta}(\vk) = (\mathbf{d}_\vk\cdot\underline{\sigma})\i\underline{\sigma_y}$, a surprising result appears. Diagonalizing Eq. (\ref{eq:H}) now yields the eigenvalues
\begin{align}\label{eq:pwave}
\varepsilon = \eta v_F|\vk| -\beta\sqrt{\mu^2+|\Delta(\vk)|^2}.
\end{align}
when we for concreteness consider a triplet state $\mathbf{d}_\vk = \Delta(\vk)\hat{\mathbf{z}}$. Remarkably, the superconducting order parameter simply renormalizes the chemical potential and the \textit{excitations remain gapless}. Several anomalous properties can be derived from the dispersion Eq. (\ref{eq:pwave}). By evaluating the corresponding wavefunction, one may conclude that Andreev reflection is strongly suppressed at the interface to a non-superconducting region since there is no gap in the charge excitation spectrum that can retroreflect a hole quasiparticle. We have checked that for any triplet symmetry the anomalous dispersion Eq. (\ref{eq:pwave}) is obtained. Thus, the results for singlet and triplet pairing differ qualitatively in a fundamental way, as the excitations are gapped in the former case whereas they remain ungapped in the latter case. One may speculate whether a superconducting state is supported at all in the spin-triplet case due to the apparent lack of a gap which offers a net condensation energy. The unusual behavior of the spin-triplet symmetry appears to be a direct result of the band-structure in the TI, where the spin couples directly to momentum through the term $\underline{\sigma}\cdot\mathbf{k}$ in the Hamiltonian. Since the spin will be parallell to the momentum, one may note that pairing between equal spins (triplet pairing) at $\vk$ and $-\vk$ is not possible. Note that this is distinct from the case of graphene, where the operator $\boldsymbol{\sigma}$ does not represent physical spin, but rather a pseudospin index related to the sublattices \cite{katsnelson_nphys_06, uchoa_prl_07}.

\begin{figure}[t!]
\centering
\resizebox{0.4\textwidth}{!}{
\includegraphics{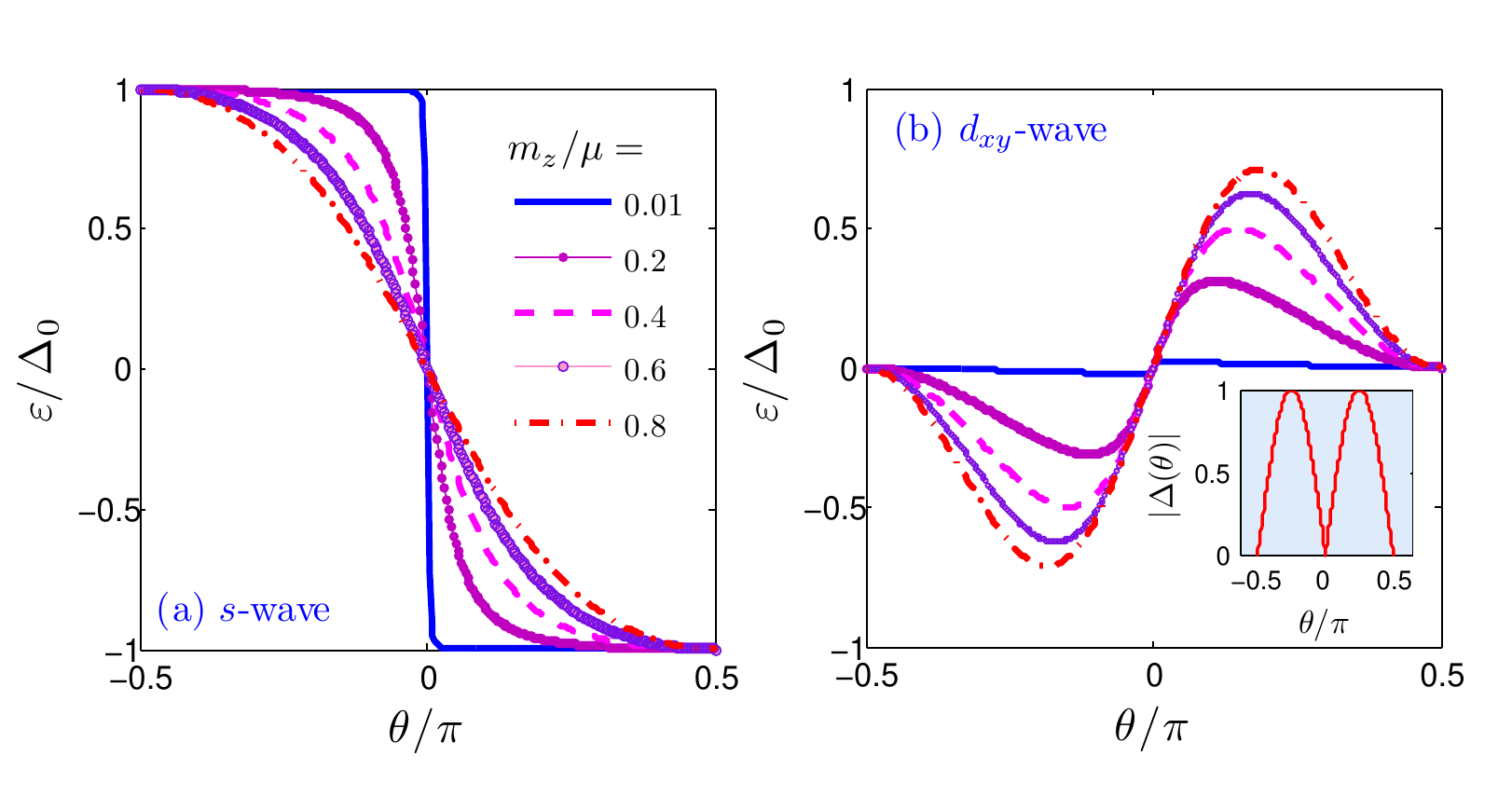}}
\caption{(Color online) Plot of the dispersion for the bound-state energies for several values of the Zeeman field $m_z$. As $m_z\to0$, one obtains $|\varepsilon|\to\Delta$ in the $s$-wave case and $|\varepsilon|\to0$ in the $d_{xy}$-wave case. We have set $\mu_S/\Delta_0=100 \gg \mu_N/\Delta_0$. The energy dispersion changes sign when $m_z$ changes sign.}
\label{fig:dispersion} 
\end{figure}

In order to investigate how Andreev reflection is influenced by unconventional pairing in the environment of a TI, we turn our attention to spin-singlet pairing and first consider the simplest experimental hybrid structure that can probe this phenomenon: a normal metal$\mid$superconductor (N$\mid$S) junction. The scattering coefficient for Andreev reflection is obtained via setting up wavefunctions and connecting them at the interface. In the N region, we have $\psi_N \propto [1,\e{\i\theta},0,0] + r_e[1,-\e{-\i\theta},0,0] + r_h[0,0,1,-\e{-\i\theta}]$ at $x=0$, while $\psi_S \propto t_e[\e{\i\beta}, \e{\i(\beta+\theta')}, -\e{\i(\theta'-\gamma_+)},\e{-\i\gamma_+}] + t_h[1,-\e{-\i\theta'}, \e{\i(\beta-\theta'-\gamma_-)},\e{\i(\beta-\gamma_-)}]$. Note that we have taken into account the possibility of anisotropic pairing such as $d$-wave, and consequently defined $\e{\i\beta} = u_+/u_-$, $u_\pm = \sqrt{\frac{1}{2}(1 \pm \sqrt{\varepsilon^2-|\Delta(\theta')|^2}/E)}$, and $\e{\i\gamma_\pm} = \Delta(\theta_\pm)/|\Delta(\theta_\pm)|$, $\theta_+=\theta', \theta_-=\pi-\theta'$. A difference in doping level between the N and S regions is accounted for by $\mu_N\sin\theta = \mu_S\sin\theta'$, since in an experimental situation the S region is often heavily doped ($\mu_S\gg\mu_N$), in which case one may set $\theta'=0$. In this case, we recover the bound-state solution $\varepsilon=0$ for $d_{xy}$-wave pairing which is manifested as a ZBCP \cite{hu_tanaka}.
\par
We now provide an argument for why the zero-energy bound-state appearing in the $d_{xy}$-wave case is a Majorana fermion, in contrast to the zero-energy states realized in the topologically trivial high-$T_c$ cuprates. The crucial factor here is the spin-degeneracy of the Fermi surface in the latter case, whereas for a TI this degeneracy is lifted. In both cases, the $4\times4$ BdG Hamiltonian $\hat{H}$ satisfies a particle-hole symmetry $\Theta\hat{H}(\vk)\Theta = -\hat{H}^*(-\vk)$, where $\Theta = \begin{pmatrix} \underline{0} & \underline{1}\\ \underline{1} & \underline{0} \\ \end{pmatrix}$ \cite{sato_prb_09}. From this property, one may prove that if $\psi_\varepsilon = [u_1(\vk),u_2(\vk),v_1(\vk),v_2(\vk)]$ is an eigenfunction for the eigenvalue $\varepsilon$, then $\Theta\psi_\varepsilon(-\vk)^* = \psi_{-\varepsilon}(\vk) = [v_1^*(-\vk),v_2^*(-\vk),u_1^*(-\vk),u_2^*(-\vk)]$ is an eigenfunction for $(-\varepsilon)$. For a zero-energy bound state $\varepsilon=0$, one must have $\psi_\varepsilon = \psi_{-\varepsilon}$, leading to internal symmetry relations between the coherence factors such as $u_1(\vk)=v_1^*(-\vk)$. The Bogoliubov quasiparticle creation operator for this state is constructed in the usual way as $\gamma^\dag(\vk) = u_1(\vk) c_\uparrow^\dag(\vk) + u_2(\vk) c_\downarrow^\dag(\vk) + v_1(\vk) c_\uparrow(-\vk) + v_2(\vk) c_\downarrow(-\vk)$. Thus, we see that the Majorana criterium $\gamma(\vk) = \gamma^\dag(-\vk)$ is satisfied. Now, the distinction between the zero-energy state in the cuprates and the present context of a TI is precisely the spin-degeneracy which allows one to split up the $4\times4$ BdG equations to two separate $2\times2$ equations per spin. Due to the band-structure on the surface of a TI, the $\varepsilon=0$ solution is not spin-degenerate and we obtain only one zero-energy mode. As pointed out in Ref. \cite{sato_prb_09}, this guarantees the Majorana nature of the fermion. We reemphasize that this is different from topologically trivial N$\mid$$d_{xy}$-wave junctions, where the zero-energy solutions are spin-degenerate, i.e. "double Majorana" modes.

Recent work has demonstrated how Majorana bound-states are induced in the presence of a Zeeman-field when contacted to a $s$-wave superconductor \cite{fu_prl_08, tanaka_prl_09}. We now wish to investigate this phenomenon when the superconducting order parameter is unconventional, i.e. we consider a N$\mid$FI$\mid$$d$-wave junction. In the $d_{xy}$-wave case, the spin-singlet order parameter reads $\underline{\Delta}(\theta) = \Delta_0\cos(2\theta-\pi/2)\i\underline{\sigma_y}$ which normally supports zero-energy states. Setting up the scattering wavefunctions and utilizing appropriate boundary conditions, one is able to extract the reflection and transmission coefficients. The chemical potential $\mu_S=\mu_N=\mu$ is assumed to satisfy $\mu\gg\Delta$ in order to accomodate proximity-induced superconductivity, except in the FI region where $\mu_{FI}=0$. Proper gating of the different regions depicted in Fig. \ref{fig:model} allow for control over the local chemical potential. Let us now define the quantities $\nu = v_F(k_y-\kappa)/m_z$ and $\kappa = \sqrt{(v_Fk_y)^2+m_z^2}/v_F$.
We then arrive at the following general expression for the bound-state energy in the limit of vanishing normal-state conductance $\sigma_\text{N}$, i.e. $L\to\infty$ where $L$ is the width of the FI region: $\varepsilon = |\Delta(\theta)|\text{sgn}\{\mathcal{C}\}/\sqrt{1 + \mathcal{C}^2},\; \mathcal{C} = \tan[\text{ln}(\zeta \mathcal{A}_-/\mathcal{A}_+)/2\i]$, $\mathcal{A}_\pm = \sin(2\theta+\delta)+\sin(\delta)\pm[\sin(2\delta+\theta)+\sin(\theta)].$
Here, we have defined $\delta=-\i\ln(\nu/\i)$ and $\zeta=-1$ in the $s$- and $d_{x^2-y^2}$-wave case while $\zeta=+1$ in the $d_{xy}$-wave case. In the $s$-wave case, this expression agrees with the very recent finding of Ref. \cite{tanaka_prl_09}. In the $d_{xy}$-wave case, a zero-energy solution exists in the absence of magnetization $m_z$, indicating the presence of midgap Andreev bound states. To explore how the magnetization influences the bound-state dispersion, we plot in Fig. \ref{fig:dispersion} the bound-state energy for both $s$-wave and $d_{xy}$-wave pairing for several choices of $m_z$. In the $s$-wave case, the bound-states have a dispersion only near $\theta=0$ when the Zeeman field is small, $|m_z|\ll\mu$. A zero-energy solution is seen to be allowed for normal incidence as long as $m_z$ is finite. In the $d_{xy}$-wave case, the dispersion is also very small for $|m_z|\ll\mu$, but in this case it lies almost at $\varepsilon=0$. Increasing $m_z$ ($m_z>0$) in the $d$-wave case has the important effect of accomodating finite-energy bound-states when moving away from normal incidence. It should also be noted that the chirality of the bound-states are determined by $\text{sgn}\{m_z\}$ in both the $s$-wave and $d$-wave case. To see this, note that $m_z\to(-m_z)$ leads to $\delta\to\delta+\pi$. Due to the symmetry relations $A_\pm(\delta+\pi) = -A_\mp(\delta)$ and ln$(A_-/A_+)$ = -ln$(A_+/A_-)$, it follows from the definition of $\mathcal{C}$ that $\text{sgn}\{\mathcal{C}\} \propto \text{sgn}\{m_z\}$.

\begin{figure}[b!]
\centering
\resizebox{0.4\textwidth}{!}{
\includegraphics{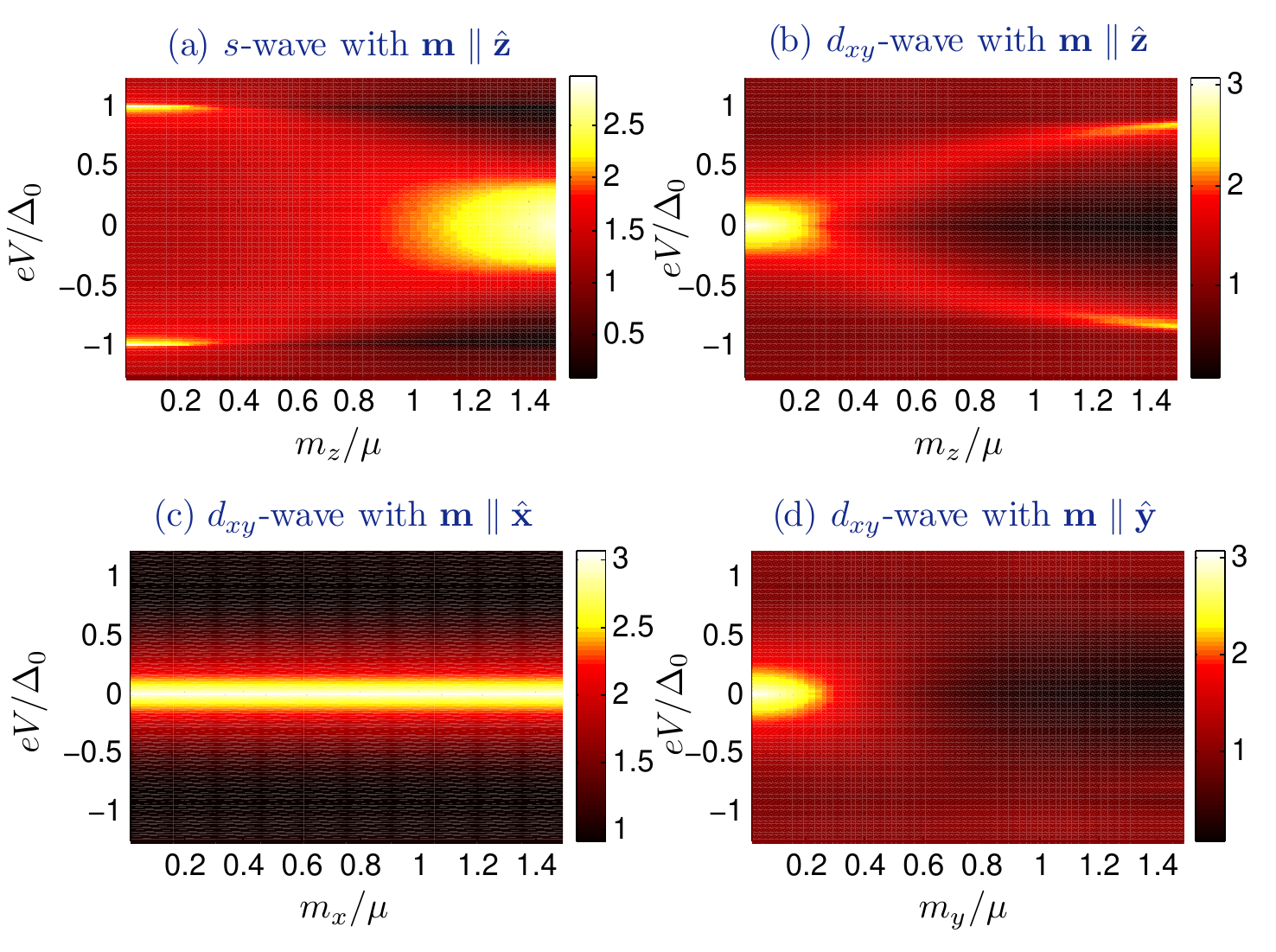}}
\caption{(Color online) Plot of the tunneling conductance $G/G_0$ for an N$\mid$FI$\mid$S junction in the $s$-wave and $d_{xy}$-wave case. We have set $\mu L/v_F=1$. 
}
\label{fig:conductance} 
\end{figure}

The experimental signature of Majorana fermions manifested as bound-states would be a characteristic behavior of the tunneling conductance, as we now show. The normalized conductance $G/G_0$ may be evaluated by defining $G = \int^{\pi/2}_{-\pi/2} \text{d}\theta\cos\theta[1 + |r_h(\theta)|^2-|r_e(\theta)|^2]$, where $r_e$ and $r_h$ denote the normal and Andreev reflection scattering coefficients, respectively, while we choose $G_0=G(|eV|\gg\Delta_0)$ as usually done in experiments. In Fig. \ref{fig:conductance}, we plot the conductance for an N$\mid$FI$\mid$S junction and compare the $s$-wave case against the $d_{xy}$-wave case. The latter is normally expected to produce the well-known ZBCP due to midgap resonant states \cite{hu_tanaka}. One of the main aims of this work is to investigate if this hallmark of the $d_{xy}$-wave state survives in the present case of a TI.

As seen in Fig. \ref{fig:conductance}(a), the conductance for the $s$-wave case displays two coherence peaks at $\varepsilon=\Delta_0$ as usual when $m_z\to 0$. Upon increasing $m_z$, the appearance of zero-energy states are manifested by a large enhancement of the zero-bias conductance. Therefore, the two finite-energy peaks are merged into one zero-energy resonance when the Zeeman field $m_z$ increases. We also note that the $d_{x^2-y^2}$-wave case is qualitatively similar to the $s$-wave case in Fig. \ref{fig:conductance}(a). Consider now the $d_{xy}$-wave case in (b), where a zero-bias peak is present when $m_z\to0$, in agreement with our previous analytical finding. However, the evolution of the conductance spectra are now opposite to the $s$-wave case upon increasing $m_z$: the zero-bias peak is split into two finite-energy resonances. In effect, this means that the characteristic features in the conductance spectra of $s$-wave and $d$-wave superconductors \textit{can be completely reversed} by introducing a Zeeman field in the TI.

Due to the coupling between spin and momentum in the band structure of the surface of a TI, it is interesting to check whether the direction of the magnetization influences the conductance spectra. In a topologically trivial N$\mid$FI$\mid$$d_{xy}$-wave junction, one can prove analytically that the conductance is invariant with respect to the direction of the magnetization $\mathbf{m}$ of the FI layer. Increasing the exchange field in the FI region, the ZBCP splits in the conventional case \cite{kashiwaya_prb_99}, similarly to Fig. \ref{fig:conductance}(b). We here show that in complete contrast to the topologically trivial case, the conductance now features a strong dependence on the magnetization orientation. We consider a magnetization in the $\hat{\mathbf{x}}$- and $\hat{\mathbf{y}}$-direction in Fig. \ref{fig:conductance}(c) and (d), respectively. It is seen that depending on the magnetization orientation, the conductance features three qualitatively different types of behavior. For $\mathbf{m}\parallel\hat{\mathbf{x}}$, $G/G_0$ is invariant upon increasing $m_x$. For $\mathbf{m}\parallel\hat{\mathbf{y}}$, the ZBCP vanishes upon increasing $m_y$. For $\mathbf{m}\parallel\hat{\mathbf{z}}$, the ZBCP is split upon increasing $m_z$. This strong sensitivity to the direction of $\mathbf{m}$ is a new feature compared the topologically trivial case which pertains directly to the anomalous band-structure of the TI. It may be understood by noting that $m_{y}$ shifts the Fermi surface
while $m_{z}$ opens the energy gap in the FI region. This places strong restrictions on how the wavefunction in the S region connects to the FI. For sufficiently large $m_y$, the Fermi surface is shifted in such a fashion
that there are no angles of incidence where surface-bound states can be
formed any more. We note that an inclusion of the orbital effect due to the vector potential $\mathbf{A}$ simply would add a component to the magnetization vector as a result of the linear energy-momentum dispersion.

The predicted results in this work can be tested experimentally by fabricating a hybrid structure such as the one shown in Fig. \ref{fig:model}. In terms of actual materials, EuO or EuS might be suitable as ferromagnetic insulators in this context \cite{tedrow_prl_86}. For the $d$-wave superconductor, a high-$T_c$ cuprate such as YBCO would be appropriate. Due to the lattice mismatch between the host proximity materials and the TI, the induced superconducting order parameter $\Delta_0$ can be expected to be substantially reduced in magnitude on the surface of the TI, typically in the range $0.1-1$ meV. Concerning the length of the sample, we have set $\mu L/v_F=1$. Estimating $v_F\simeq 5.0\times10^5$ m/s and $\mu\simeq80$ meV \cite{zhang_nphys_09} , this corresponds to $L\simeq 40$ nm which should be experimentally feasible.

In summary, we have considered the interplay between magnetic order and unconventional superconducting pairing on the surface of a topological insulator. We find that the charge excitation spectrum is rendered gapless for any spin-triplet state, such that both bound-states and Andreev reflection are strongly suppressed. For spin-singlet pairing, we find that the zero-energy surface states in the $d_{xy}$-wave case are now Majorana fermions, in contrast to the case of the topologically trivial high-$T_c$ cuprates. We have studied how Andreev-bound states and Majorana fermions are influenced by the internal phase of the superconducting order parameter, and find that the ZBCP being the hallmark of the $d_{xy}$-wave state is qualitatively strongly modified in the present context. In particular, it is highly sensitive to the magnetization orientation, in contrast to the topologically trivial case. Our findings can be directly tested through tunneling spectroscopy measurements, and we have estimated the magnitude of the necessary experimental quantities.

\textit{Acknowledgments.} 
M. Sato and M. Cuoco are thanked for very useful discussions. J.L. and A.S. were supported by the Research Council of Norway, 
Grants No. 158518/431 and No. 158547/431 (NANOMAT), and Grant No. 167498/V30 (STORFORSK). T.Y. acknowledges support by JSPS.

\end{document}